# Effects of dopants in $InO_x$-based amorphous oxide semiconductors for thin-film transistor applications

Shinya Aikawa,[1,a)] Toshihide Nabatame,[2] Kazuhito Tsukagoshi[1,b)]

[1] International Center for Materials Nanoarchitectonics (WPI-MANA), National Institute for Materials Science (NIMS), Tsukuba, Ibaraki 305-0044, Japan

[2] MANA Foundry and MANA Advanced Device Materials Group, National Institute for Materials Science (NIMS), Tsukuba, Ibaraki 305-0044, Japan

Corresponding authors:  [a)] E-mail: AIKAWA.Shinya@nims.go.jp,
[b)] E-mail: TSUKAGOSHI.Kazuhito@nims.go.jp

**ABSTRACT**

Amorphous metal oxide thin-film transistors (TFT) are fabricated using $InO_x$-based semiconductors doped with $TiO_2$, $WO_3$ or $SiO_2$. Although density of dopant is low in the film, change in the electrical properties showed strong dependence on the dopant species. We found that the dependence could be reasonably explained by the bond-dissociation energy. By incorporating the dopant with higher bond-dissociation energy, the film becomes less sensitive to oxygen partial pressure used during sputtering deposition and remains electrically stable to thermal annealing treatment. The concept of bond-dissociation energy can contribute to the realization of more stable metal oxide TFTs for flat panel displays.

Recently, $InO_x$-based metal oxide semiconductors, such as pure In-O,[1,2] In-Ga-O,[3,4] In-Sn-O (ITO),[5,6] In-Zn-O (IZO),[7,8] In-W-O,[9] In-Ga-Zn-O (IGZO),[10,11] and Ta-In-Zn-O,[12] are widely studied for thin-film transistor (TFT) applications because they exhibit high electron mobility due to band conduction originating from an edge-sharing polyhedral structure.[13,14] In particular, the amorphous IGZO TFT has far superior electrical properties than those of conventional amorphous Si TFTs, which has contributed to the development of next-generation high-definition display panels based on liquid-crystal and organic light-emitting diodes (OLEDs).[15] Flexible and transparent TFTs can be fabricated on plastic substrates if metal oxide materials are used for all components, including the active and dielectric layers, and the electrodes.[11] It is widely known, however, that an amorphous IGZO TFT is still difficult to adjust optimal TFT performances in terms of device stability.[16]

We have investigated the In-W-O system to develop W-doped $InO_x$ semiconductors.[9] This system is completely free from both Ga and Zn and TFTs that employ this type of sputtered thin-film that exhibits superior electrical properties due to the very flat and smooth surface of the amorphous structure. However, the role of W atoms in the film has not yet been fully clarified, so that W doping may possibly not be the best dopant choice for $InO_x$ semiconductors. Therefore, to understand the role of small amounts of dopants in $InO_x$ semiconductors, $InO_x$-based films with W, Ti or Si dopants were compared. Here, we propose the role of the dopant in an $InO_x$-based semiconductor film in terms of the bond-dissociation energy with oxygen atoms. Consideration of the bond-dissociation energy is demonstrated as a useful concept for the selection of appropriate dopants for $InO_x$-based metal oxide semiconductors.

Oxide TFTs were fabricated on a heavily doped p-type Si substrate with a 285 nm thick thermal $SiO_2$ layer. The substrate was first cleaned using a standard process of ultrasonication in acetone and isopropyl alcohol, followed by UV-plasma exposure. A patterned $InO_x$-based semiconductor film with a thickness of 10 nm was then deposited through a stencil shadow mask at room temperature under various oxygen partial pressure conditions using DC magnetron sputtering. The sputtering power was fixed at 50 W and the targets of In-Ti-O, In-W-O, and In-Si-O contained 1 wt% of $TiO_2$, $WO_3$, or $SiO_2$, respectively, in $In_2O_3$. There was no inclusion of Zn as a target compression mixture as previously employed in our initial work on the In-W-O system.[9] Source and drain electrodes (40 nm thick Au) were formed by vacuum evaporation through a stencil shadow mask. The device was then annealed at 150 °C in the ambient air. The underlying Si substrate was employed as a back-gate electrode. The device configuration without any passivation layer is shown in Fig. 1(a). Film characterization using X-ray diffraction and atomic force microscopy indicate that all of the films remained in the amorphous state. Thus, the low temperature annealing at 150 °C, which is better for application to the OLED display process, is still insufficient for crystallization of these materials.

Figures 1(b)-(g) show the typical electrical properties of the fabricated TFTs annealed at 150 °C. All measurements were conducted at room temperature in the dark under the ambient atmosphere using a precision semiconductor parameter analyzer (Agilent 4156C). Figures 1(b), (d), and (f) indicate the output characteristics ($I_D$-$V_{DS}$) for the three TFTs with different dopants. The transfer characteristics ($I_D$-$V_{GS}$) in the saturation region ($V_{DS}$ = 30 V) for In-Ti-O [Fig. 1(c)], In-W-O [Fig. 1(e)], and In-Si-O [Fig. 1(g)] have negligibly small hystereses and high on/off current ratios ($I_{ON}/I_{OFF}$). The observed TFT properties are summarized in Table I. The saturation field-effect mobilities ($\mu_{sat}$) extracted using $\mu_{sat} = (\partial\sqrt{I_D}/\partial V_{GS})^2(2L/W)(1/C_i)$ were 32, 30, and 17 cm$^2$ V$^{-1}$ s$^{-1}$ for In-Ti-O, In-W-O, and In-Si-O, respectively. For the extraction, the channel length $L$ (350 μm) and width $W$ (1000 μm), and gate capacitance per unit area $C_i$ (1.21×10$^{-8}$ F/cm$^2$ based on a dielectric constant of 3.9 for SiO$_2$) were used. The field-effect mobility for In-Si-O is slightly suppressed, which may be due to the incorporation of Si introduced into InO$_x$ to stabilize the semiconductor properties. However, $\mu_{sat}$ for In-Si-O can be improved by optimizing the O$_2$ partial pressure used during sputter deposition. Further improvement can be expected by modifying the contact electrode material[17] and by utilizing Ar plasma treatment[18] or excimer laser irradiation.[19]

Figure 2 shows the electrical conductivity of the as-deposited films as a function of the oxygen partial pressure used during sputter deposition. Each value was extracted from the linear region of the output characteristics at $V_{GS}$ = 30 V. Similar to other metal oxide semiconductors such as IGZO[20,21] and ZnO,[22] the electrical conductivity of the InO$_x$-based films is dependent on the O$_2$ partial pressure because the carrier concentration is strongly dependent on O$_2$ partial pressure employed during deposition.[23,24] Although the dopant content is only 1 wt%, the change in the electrical conductivity is very pronounced, depending on the dopant species. In the case of In-Ti-O, the electrical conductivity changes significantly with the oxygen partial pressure for deposition; however, this is not the case for In-Si-O. The In-Si-O system exhibits the most stable electrical conductivity among the three types of film. Thus, it is considered that the slope of the electrical conductivity as a function of the oxygen partial pressure could be related to the bond-dissociation energy. The bond-dissociation energy is defined as the strength of the chemical bond determined when the diatomic species is decomposed into individual atoms. If the energy of the metal to oxygen bond (X-O, where X is the dopant material) is low, then oxygen can be easily released, which results in an increase of the carrier density.[24,25] In contrast, if a dopant with a high bond-dissociation energy (to oxygen atom) is introduced into InO$_x$, the electrical properties of the oxide film would be expected to be more stable over a range of oxygen partial pressure used during deposition. According to the results presented in Fig. 2, the slopes become flatter in the order of In-Ti-O, In-W-O, and In-Si-O, which corresponds to the order of the bond-dissociation energies; 666.5, 720, and 799.6 kJ/mol for Ti-O,[26] for W-O,[27] and Si-O,[28] respectively. Although molar ratio of each sputtering targets are different as shown in Table I, the slopes are dominated by the bond-dissociation energy.

The stability of the electrical conductivity was dependent on the thermal annealing treatment, as shown in Fig. 3, which shows a comparison of the electrical conductivity for In-Ti-O and In-Si-O with different annealing treatment times. The slope for In-Ti-O changes significantly with increased annealing time, while that for In-Si-O does not change. This difference can be understood with respect to the weaker bond-dissociation energy of Ti-O than that of Si-O. On the other hand, the electrical conductivity of In-Si-O is less sensitive to the annealing treatment over the entire $O_2$ partial pressure range.

The change in conduction is also reflected in the transfer characteristics [Figs. 4(a) and (b)]. The shape of the *I-V* curves for In-Ti-O changes after thermal annealing for 5 min. The off-state region in the In-Ti-O TFT cannot be observed in the range between −20 and 40 V, even after annealing at 150 °C for 10 min. In contrast, the semiconductive properties of In-Si-O remain constant, although the annealing time is relatively longer. An increased annealing time for the In-Ti-O TFT results in an upsurge of the subthreshold slope (*ss*) and a decrease in the $I_{ON}/I_{OFF}$ ratio, whereas these values are stable for the In-Si-O TFT [Fig. 4(c)]. The changes in the conductivity and TFT properties with thermal annealing treatment strongly support that the bond-dissociation energy can be used to determine the role of the dopant in the $InO_x$-based semiconductor system.

Finally, we discuss conventional InO$_x$-based metal oxide semiconductors in terms of the bond-dissociation energy. Based on an analogical comparison of the mobility difference between IGZO and IZO, it is reasonable that the lower mobility of IGZO could be due to the incorporation of Ga, as suggested by Kamiya *et al.*, which is important to obtain stable electrical properties for a TFT,[14] because the bond-dissociation energy of Ga-O is 374 kJ/mol, which is slightly higher than In-O.[29] Thus, if a large amount of Ga atoms is introduced into IZO, then the electrical properties become more stable. In the present work, improvement of the electrical stability was examined by utilizing dopants with higher oxygen bond-dissociation energies (Ti-O = 666.5; W-O = 720; Si-O = 799.6 kJ/mol). The resultant In-Si-O TFT showed higher electrical stability after the annealing process, although the dopant density was rather low (1 wt%). According to first-principle calculations,[30] electrical stability of amorphous oxide semiconductors becomes effectively controllable by incorporation of Si since formation of oxygen vacancies can be suppressed due to strong binding of Si-O. In terms of bond strength to oxygen atoms, Hf-doping is also expected similar effect to Si.[31-33] However, in order to clarify the carrier transport in the present InO$_x$-based semiconductors based on thermal activation, energy band diagram and local atomic structure, further characterizations such as low temperature *I-V* measurement,[34] Hall measurement,[35,36] and X-ray absorption spectroscopy[13,37] will be discussed in another report.

In conclusion, amorphous oxide TFTs were fabricated using InO$_x$-based semiconductors doped with Ti, W or Si atoms. These dopant atoms have different bond-dissociation energies to oxygen; therefore, the electrical properties of the films are dependent on the bond-dissociation energy of X-O in $InO_x$-based metal oxide semiconductors. The selection of optimal dopant materials based on the concept of bond-dissociation energy is expected to expand the processing window of sputter deposition conditions, especially the $O_2$ partial pressure. This trend could then be exploited to prepare more stable oxide films, such as the example of In-Si-O in the present study, which included 1 wt% of $SiO_2$ in an $In_2O_3$ target. The concept of bond-dissociation energy presented here could be very important for the development of post-IGZO metal oxide semiconductor materials for next-generation flat panel displays.

**Acknowledgements**

The authors would like to thank Prof. S. Maruyama and Prof. S. Chiashi (The University of Tokyo) and Dr. Y. Xu, Dr. M. Shimizu and Dr. T Kizu (WPI-MANA) for fruitful discussions. The authors also acknowledge K. Yanagisawa (RIKEN) for support in the preparation of the oxide films and for valuable discussions. The authors thank Sumitomo Metal Mining Co., Ltd. for supplying the sputtering targets ($TiO_2$-doped $In_2O_3$ and $WO_3$-doped $In_2O_3$).

**Table**

Table I. Comparison of typical TFT properties estimated from the transfer characteristics presented in Fig. 1. $I_{ON}/I_{OFF}$ is defined as the maximum to minimum $I_D$ ratio in the graph ($-20 \leq V_{GS} \leq 40$). The subthreshold slope and threshold voltage are denoted as $ss$ and $V_{th}$, respectively. The turn-on voltage $V_{on}$, is defined as $V_{GS}$ at which $I_D$ starts to increase.

|  | $I_{ON}/I_{OFF}$ | $ss$ (V/dec) | $V_{th}$ (V) | $V_{ON}$ (V) | $\mu_{sat}$ (cm$^2$ V$^{-1}$ s$^{-1}$) | Molar ratio of the sputtering target |
|---|---|---|---|---|---|---|
| In-Ti-O | $9.4 \times 10^9$ | 0.30 | 1.0 | $-4$ | 32 | $In_2O3 : TiO_2$ = 1 : 0.035 |
| In-W-O | $5.2 \times 10^9$ | 0.46 | 0.9 | $-6.3$ | 30 | $In_2O3 : WO_3$ = 1 : 0.012 |
| In-Si-O | $4.8 \times 10^9$ | 0.29 | 5.3 | 0 | 17 | $In_2O3 : SiO_2$ = 1 : 0.047 |

**Figure captions**

Fig. 1. (a) Optical micrograph and schematic illustration of the fabricated device. Typical output and transfer characteristics of fabricated TFTs: (b, c) In-Ti-O, (d, e) In-W-O, and (f, g) In-Si-O. The devices were annealed at 150 °C in air for optimized periods; the optimal annealing times for In-Ti-O, In-W-O, and In-Si-O were typically 5, 10, and 15 min, respectively. The TFTs were measured at room temperature in the dark under ambient atmosphere.

Fig. 2. Electrical conductivity of as-deposited films as a function of the oxygen partial pressure used during sputtering deposition. The film conductivity was extracted from the linear region of the output characteristics at $V_{GS}$ = 30 V.

Fig. 3. Changes in the electrical conductivity of In-Si-O (circles) and In-Ti-O (triangles) after annealing at 150 °C in air. The annealing time was varied between 5 and 15 min.

Fig. 4. Transfer characteristics for (a) In-Ti-O and (b) In-Si-O TFTs after annealing at 150 °C in ambient air for 5-15 min. (c) Extracted TFT parameters (field-effect mobility in the saturation region, subthreshold slope (ss) and on-off current ratio) for In-Ti-O and In-Si-O. $I_{ON}/I_{OFF}$ was defined as the maximum to minimum drain current ratio in graphs (a) and (b).

**Fig. 1**

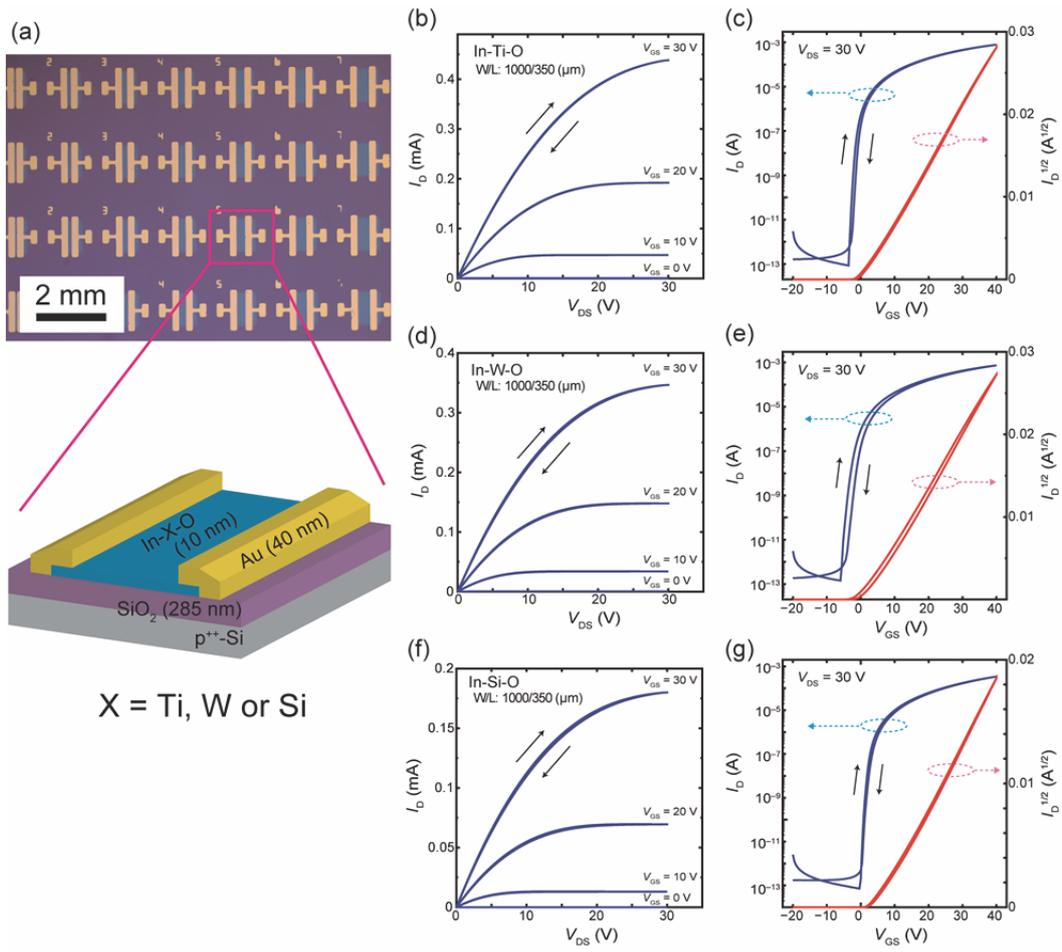

**Fig. 2**

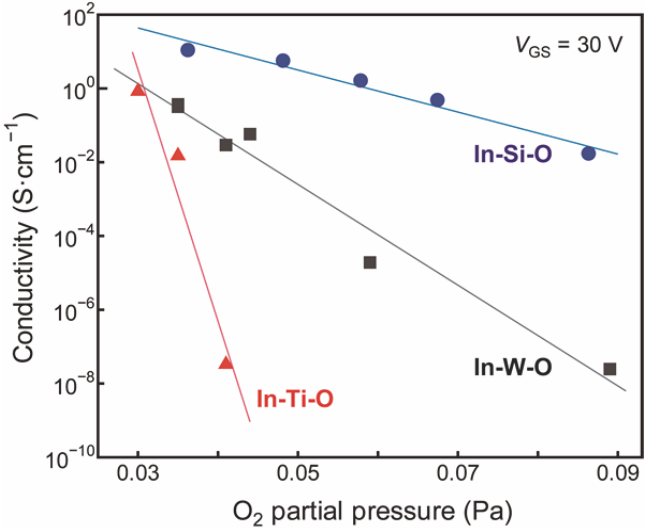

**Fig. 3**

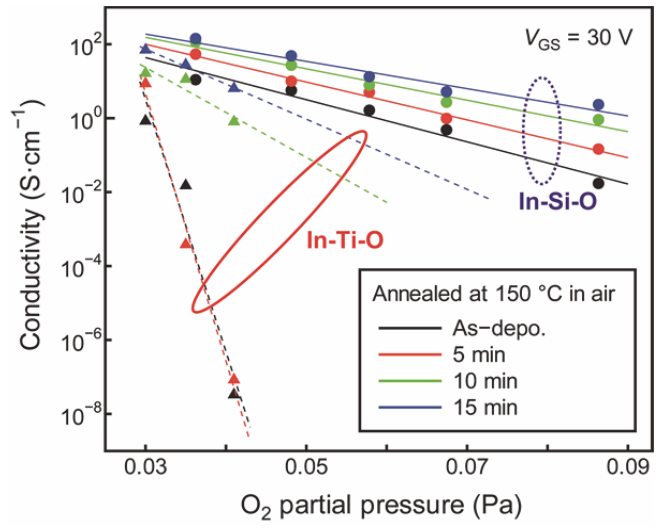

**Fig. 4**

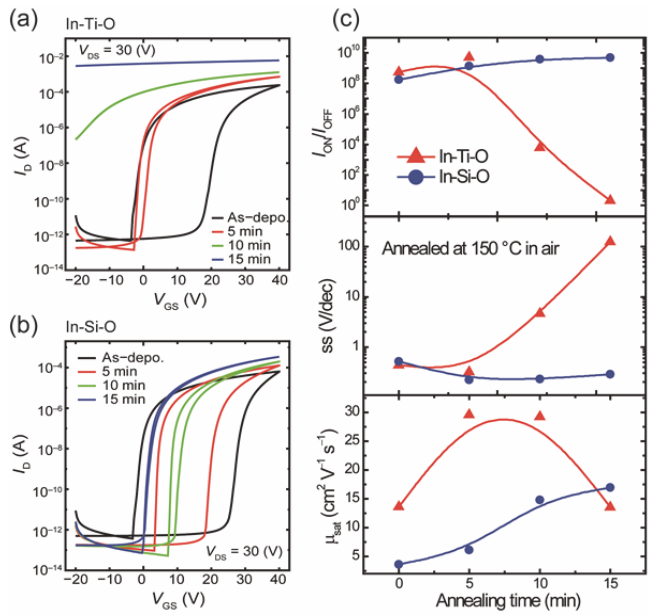